\documentclass[runningheads]{svmult}

\usepackage{makeidx}   % allows index generation
\usepackage{graphicx}  % standard LaTeX graphics tool
\usepackage{subeqnar}  % subnumbers individual equations
\usepackage{multicol}  % used for the two-column index
\usepackage{physprbb}  % modified textarea for proceedings,
\makeindex             % used for the subject index
\begin{document}
\title*{Dynamic Model Atmospheres of Cool Giants}
%
%
%%  \toctitle{Focusing of a Parallel Beam to Form a Point
%%  \protect\newline in the Particle Deflection Plane}
% allows explicit linebreak for the table of content
%
%
%%  \titlerunning{Dynamic Model Atmospheres of Cool Giants}
% allows abbreviation of title, if the full title is too long
% to fit in the running head
%
\author{Susanne H{\"o}fner\inst{1,2}
\and Rita Gautschy-Loidl\inst{4}
\and Bernhard Aringer\inst{3}
\and Walter Nowotny\inst{3}
\and Josef Hron\inst{3}
\and Bernd Freytag\inst{1}}
\authorrunning{S. H{\"o}fner et al.}
% if there are more than two authors,
% please abbreviate author list for running head
%
%
\institute{Dept. of Astronomy \& Space Physics, Box 515, SE-75120 Uppsala, Sweden
\and NORDITA, Blegdamsvej 17, DK-2100 Copenhagen, Denmark
\and Inst. f. Astronomie, T{\"u}rkenschanzstr. 17, A-1180 Vienna, Austria
\and Froburgstr. 43, CH-4052 Basel, Switzerland}

\maketitle              % typesets the title of the contribution

\begin{abstract}
Cool giant stars are highly dynamical objects, and complex 
micro-physical processes play an important role in their extended 
atmospheres and winds. The interpretation of observations, and 
in particular of high-resolution IR spectra, requires realistic 
self-consistent model atmospheres. Current dynamical models 
include rather detailed micro-physics, and the resulting synthetic 
spectra compare reasonably well with observations. A transition from 
qualitative to quantitative modelling is taking place at present.
We give an overview of existing dynamical model atmospheres for 
AGB stars, discussing recent advances and current trends in modelling.
When comparing synthetic spectra and other observable properties 
resulting from dynamical models with observations
we focus on the near- and mid-IR wavelength range.
\end{abstract}

\section{Introduction}

Cool giant stars are prime targets for high-resolution 
infrared spectroscopy:
these stars are luminous, emit most of their radiation at near- and
mid-infrared wavelengths, and they have highly dynamical atmospheres due 
to convection and pulsation, leading to strong spectral variations. 
The coolest and most luminous of these objects also have slow, massive 
stellar winds. They represent an important but not so well-understood 
stage in the evolution of low- and intermediate mass stars, 
and they contribute significantly to the chemical enrichment 
of the interstellar medium. 
These are good reasons for studying cool giants but 
the interpretation of the observed spectra requires 
detailed and consistent model atmospheres for this type of stars.

Several of the very reasons that make pulsating asymptotic giant branch 
(AGB) stars interesting targets for IR spectroscopy, however, cause 
major challenges for realistic, self-consistent modelling of their
dynamical atmospheres and circumstellar envelopes. Many interacting 
processes have to be taken into account simultaneously: propagating shock 
waves caused by stellar pulsation modify the structures of the atmospheres 
on local and global scales, leading to strong deviations from hydrostatic 
stratification. The radiative fields are dominated by molecular opacities 
or even by dust grains forming in the cool outer layers of the atmospheres. 
Important micro-physical processes like gas-phase chemistry and dust 
formation may be severely out of equilibrium. 

Pioneering models for dynamic 
atmospheres of AGB stars date back more than two decades, but these models 
were severely restricted by the lack of sufficiently
powerful computers and micro-physical 
input data. Only recently it has become possible to calculate consistent 
dynamic models which produce reasonably realistic synthetic spectra for a 
wide range of stellar parameters. This progress is of particular interest 
in the context of current instrumental developments in IR spectroscopy.

\section{Dynamic model atmospheres}\label{s:dma}

We begin this section with a definition of the term 
'dynamic model atmospheres', in contrast to 
classical hydrostatic model atmospheres, stationary wind models, 
or pure circumstellar envelope models (which do not include 
the stellar atmosphere and wind acceleration region).
The computational domain of such 
dynamic model atmospheres is a spherical shell, 
extending from an inner boundary below the stellar photosphere
to an outer boundary which can be anywhere 
up to several tens of stellar radii above the photosphere, 
depending on whether the models have a stellar wind or not.
The stellar pulsation is usually simulated by a variable 
inner boundary ('piston models') and outflow of matter is 
permitted at the outer boundary in the case of a stellar wind.
Within this radial range, the time-dependent equations of 
hydrodynamics are solved, accounting for forces due to gravity, 
gas pressure and radiative acceleration of molecules and dust grains.
Simultaneously, a more or less sophisticated treatment of 
radiative transfer is used to determine 
energy exchange.\footnote{ 
 A general description of the relevant physical background can 
 be found, e.g., in Gustafsson \& H{\"o}fner (\cite{gh04}).}
The results of the numerical simulations are consistent spatial structures 
of velocities, densities, temperatures and other physical quantities 
as a function of time. Furthermore, these models may give 
mass loss rates, dust-to-gas ratios and other stellar wind properties.

At present, four major 'strains' of dynamical models are in use.
They share the basic ingredients given above but differ in the 
treatment of certain physical processes and the numerical methods.
Listing them by their place of origin, their special features 
(and main references) can be summarised as follows:
\begin{itemize}
\item
  The Australian models (Wood~\cite{w79}, Bessel et al.~\cite{bsw96}, 
  Hofmann et al.~\cite{hsw98})
  are strongly linked to pulsation models, featuring the most 
  consistent description of the variable inner boundary available
  at present (cf. Sect.~\ref{s:trends}).
  They have been used to derive synthetic monochromatic 
  radii and molecular line profiles to investigate the pulsation mode
  of Miras (cf. Sect.~\ref{s:spect}).
\item
  The Iowa models (Bowen~\cite{b88}; see also Willson~\cite{law00}) 
  explore the effects of non-LTE cooling on mass loss, describing 
  how thermal winds can be driven due to a 'calorisphere', i.e. 
  a region heated by shock waves and kept warm by inefficient 
  radiative cooling. Alternatively, in the case of rapid cooling
  these models use a parameterised dust opacity to produce outflows.
\item
  The Berlin models (Fleischer et al.\cite{fgs92}, 
  Winters et al.~\cite{wljhs00}, Jeong et al.~\cite{jwls03})
  introduce a detailed time-dependent description of dust formation
  which results in a non-linear interaction between gas, dust, and 
  radiation, leading to complex structures and variability.
  The models are mainly applied to determine mass loss rates and other 
  wind properties, especially for cool, dust-enshrouded stars.
\item
  The Vienna models (H{\"o}fner \& Dorfi~\cite{hd97}, 
  H{\"o}fner et al.~\cite{dma1},~\cite{dma3})
  include a similar detailed description of dust formation for 
  C-rich chemistry, and they combine time-dependent hydrodynamics with 
  frequency-dependent radiative transfer. The latter allows to 
  take the effects of molecular line blanketing properly into account,
  leading to more realistic atmospheric structures as demonstrated 
  by comparison of synthetic spectra with observations 
  (cf. Sect.~\ref{s:spect}).
\end{itemize}
An overview of dynamical models, 
including a detailed comparison of input physics, 
is given in the recent review by Woitke~(\cite{pw210}). 
The application of wind models to stellar evolution 
is discussed by Willson~(\cite{law00}).

\section{Infrared spectra: models meet observations}\label{s:spect}

When modelling the infrared spectra of cool giants, 
there are two basic philosophies. One approach, usually driven 
by new observational data, is semi-empirical modelling of specific 
observable properties of given stars. 
Examples are semi-empirical models for 
molecular line profiles which allow to deduce information about the  
velocity structure of atmospheres and winds 
(e.g. Keady et al.~\cite{khr88}, Keady \& Ridgway~\cite{kr93})
or 'composite' models of the stellar photosphere and 
circumstellar material for the interpretation of ISO spectra, 
e.g. 'slab' models of levitated warm molecular gas 
(e.g. Yamamura et al.~\cite{ydjc99}, Matsuura et al.~\cite{mycom02}).

The other method, often 
motivated by theoretical interests, is the construction of 
more generally applicable self-consistent models, and iterative 
improvement of these models by comparison with observations.
The models discussed in Sect.~\ref{s:dma} fall into this second
category: they aim at describing the time-dependent structure 
of the atmosphere of AGB stars -- and in many cases the mass loss 
through a stellar wind -- from first principles. In a second step, 
snapshots of these structures are used as input for detailed 
radiative transfer calculations, yielding synthetic spectra and 
other observable properties (colours, monochromatic radii, etc.)
which can be compared to observations.

\subsection{Consistency, all the way ?}

Considering the issue of consistency, it is, in principle,
not sufficient if the structure of the model is calculated 
in a self-consistent way. The {\it a posteriori} computation of 
observable properties has to be consistent with the 
assumptions and micro-physical data used in the dynamical model.
In the ideal case (which is often realized in classical static 
atmospheres) the calculation of synthetic spectra should just be 
a re-computation of the radiative transfer as in the original model 
but with a higher spectral resolution.

In practice, however, there are different degrees of inconsistency 
between physical assumptions made when calculating
model structures or synthetic spectra, and it depends on 
the problem under investigation what is acceptable and what is not.
For example, when studying specific spectral lines which would not make 
a noticeable contribution to the total opacity, and therefore 
not influence the global energy and momentum balance of the model,
it may be sufficient to introduce these lines only in the detailed radiative 
transfer. On the other hand, strong opacity sources in wavelength
ranges with high flux (e.g. IR bands of abundant molecules like   
water vapour) may influence the structure of the model 
considerably, and it is essential to use the same opacity data 
both in the dynamical calculation and in the computation of the 
observable properties (see contribution by Aringer et al., this volume).

An even more problematic issue is the use of grey opacities in the 
calculation of the model structures, as is often done to save computing 
time in dynamical models. This may lead to unrealistic density-temperature 
structures, a problem that has sometimes been circumvented by 
only using the density and velocity from the dynamical model, and 
then recalculating the temperature for the computation of the 
spectra (e.g. Bessel et al.~\cite{bsw96}). 

Neglecting the effects of velocity fields on spectral lines 
when evaluating the opacities during dynamical calculations
may be a borderline case of acceptable inconsistency: 
in contrast to, e.g., line-driven winds of hot stars, the 
Doppler shifts are generally small in cool giants, and the molecular 
line blanketing probably makes the effect even less pronounced.
Other sources of uncertainty in this context are non-LTE effects 
(due to shocks), 
chemical non-equilibrium, or convective motions, which are not included 
in standard dynamical model atmospheres. At present, it is difficult 
to estimate which influence these phenomena have on the structures 
of cool atmospheres and winds.

\subsection{Crucial tests for dynamical models} 

Before dynamical model atmospheres can be applied as a tool 
for the interpretation of observations, they have to be subjected 
to a number of tests which help to evaluate their consistency and 
reliability. The most simple check is, of course, a comparison of 
the hydrostatic limit case with classical model atmospheres. 
Testing of the full dynamical models, however, has to be done 
against other dynamical models and observations. In the latter 
case, it is important to pick a number of criteria which check 
different aspects of the models: 
\begin{itemize}
\item
  The global energy distribution is usually a good indicator
  for the  overall structure since different parts of 
  the spectrum are generated in different layers. 
  Therefore, photometric colours and low-resolution spectra 
  which cover a wide wavelength range 
  are important for gauging the models.
\item
  High-resolution spectra test both the structure and dynamics
  of the extended atmosphere and wind acceleration region.
  They make a quantitative determination of velocities possible
  and allow to distinguish between absorption and emission 
  components in crowded molecular bands.
\item
  Time series of observations resulting from monitoring of targets
  give insights into time-dependent processes and the global 
  dynamics which cannot be extracted from individual spectra.
  They are both the most crucial test and the most promising 
  application for dynamical models.
\end{itemize}
Another important aspect is that these various criteria 
have to be fulfilled simultaneously, i.e. a sequence of 
snapshots from one dynamical model has to fit all the 
available data for a given star, in order to claim that 
the model is consistent and realistic.

The models currently found in the literature usually only 
match part of the criteria given above, indicating that 
they are probably still too specialised on certain aspects 
and do not take all necessary processes properly 
into consideration. In the following sections we will 
look at examples, focusing on the infrared part of the 
spectrum.

\begin{figure}[p]
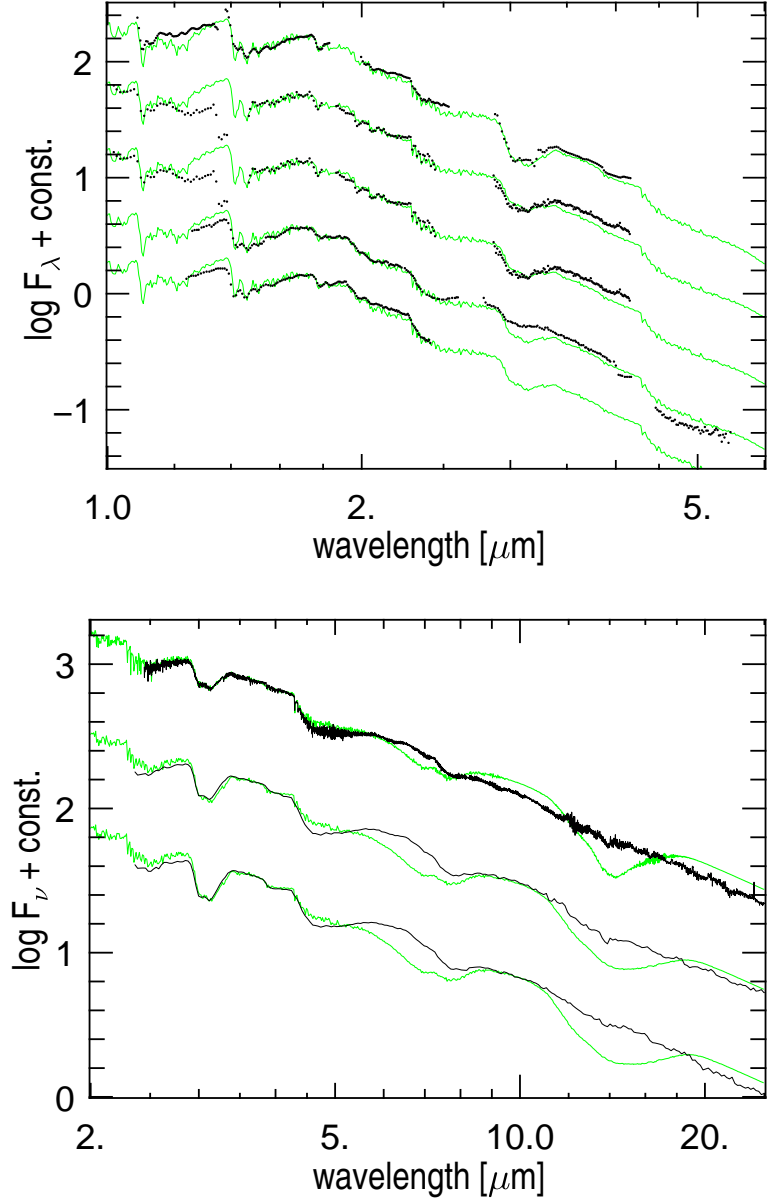

\includegraphics[width=0.9\textwidth]{hoefnerF1a.epsi}
\includegraphics[width=0.9\textwidth]{hoefnerF1b.epsi}
\caption{
  Synthetic spectra (grey) and observational data of TX Psc (black;
  upper panel: data from Joyce \cite{jo98}, 
    L\'azaro et al. \cite{lhc94} and two KAO spectra;
  lower panel: ISO SWS06 and ISO SWS01 spectra); 
  from Gautschy-Loidl et al.~(\cite{dma4}).
}\label{f:cspec1}
\end {figure}

\subsection{Low-resolution spectra - the global picture}

The spectrometers of the {\it Infrared Space Observatory} 
(ISO) opened possibilities 
to obtain a simultaneous coverage of a wide wavelength range, 
including spectral regions not accessible from the ground.
This provided a unique way of studying the global structures 
of atmospheres and inner wind regions of AGB stars and
lead to a renewed interest in the modelling of 
infrared spectra of these stars. 
After an initial phase dominated by semi-empirical modelling 
(e.g. Yamamura et al.~\cite{ydjc99}, Matsuura et al.~\cite{mycom02}),
studies based on reasonably realistic, self-consistent dynamical models 
are beginning to emerge, both for the interpretation of ISO-SWS data,
and ground-based observations.

Tej et al.~(\cite{tlsw03}) discuss synthetic optical and NIR spectra
resulting form the dynamical models of Bessell et al.~(\cite{bsw96}) 
and Hofmann et al.~(\cite{hsw98}). They find reasonable agreement 
of the overall energy distribution between 0.5 and 2.5$\,\mu$m
with spectra obtained by Lan\c{c}on \& Wood \cite{lw2000}.
They do not, however, attempt to identify a single 'best fit' 
dynamical model for each star, since the sample of dynamical 
models is small and various spectra from different models 
may reproduce the observed spectra equally well. Furthermore, 
they discuss the problem of fitting the region around 1 $\mu$m
which can possibly be attributed to missing or incorrect 
molecular opacities.

Hron et al. (\cite{Hron98}) presented a first comparison of
ISO-SWS spectra of R Scl with synthetic spectra based 
on the dynamical models of H{\"o}fner et al. (\cite{dma1}).
While these (grey) models were still too crude to 
allow for a quantitative fit, they could reproduce 
the variation of molecular features qualitatively.
Since then, the dynamical models have been improved 
by including non-grey radiative transfer, which turned 
out to be crucial for obtaining more realistic structures
(H{\"o}fner \cite{H99}, H{\"o}fner et al.~\cite{dma3}). 
The observable properties resulting from this new generation 
of dynamical models show a dramatic improvement compared 
to earlier grey models (see, e.g., Aringer et al.~\cite{Aringer02}, 
Andersen et al.~\cite{dust5}).

Recently, Gautschy-Loidl et al. (\cite{dma4}) have performed 
a systematic comparison of spectra resulting from the 
non-grey models of H{\"o}fner et al.~(\cite{dma3}) 
with observations of the C-rich AGB stars 
TX Psc, WZ Cas, V460 Cyg, T Lyr and S Cep
(ISO spectra, spectra from KAO and ground-based observations, 
narrow-band colours between 0.5 and 1.1$\,\mu$m).
In the wavelength range between 0.5 and 5$\,\mu$m,
they find good agreement between observations at different 
phases and a single model for each star.
Figure~\ref{f:cspec1} shows a comparison of 
observational data for TX Psc from various sources
with synthetic spectra based on a single dynamical model. 
The 1-5 $\mu$m range (upper panel) is well reproduced 
but there is an interesting discrepancy between 10 and 20 $\mu$m
(lower panel):
the synthetic spectra of the pulsating but dust- and windless model 
show a pronounced absorption feature around 14 $\mu$m which is not 
seen in the observations. Similar problems have been encountered 
previously when fitting hydrostatic model atmospheres to observed 
spectra of C-rich AGB stars (e.g. J{\o}rgensen et al.~\cite{jhl00}).

\begin{figure}[t]
\includegraphics[width=0.9\textwidth]{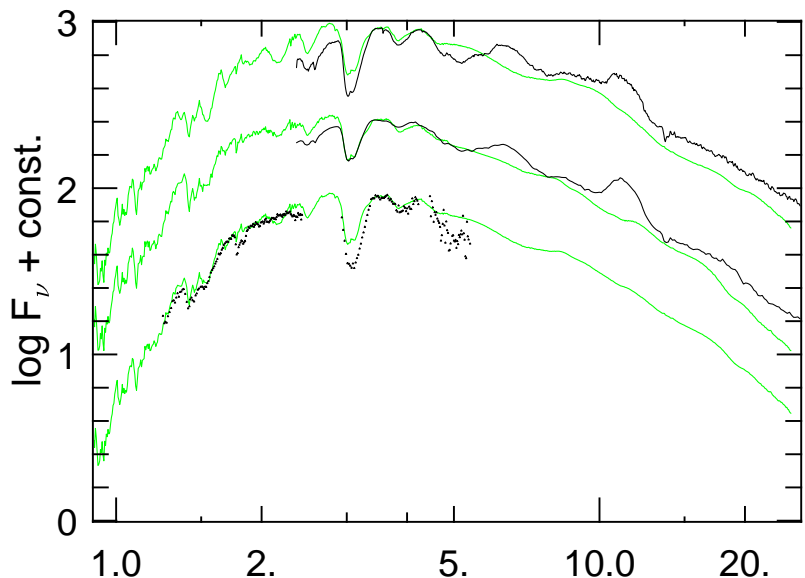}
\caption{
  Synthetic spectra (grey) and observational data of S Cep (black;
  two ISO SWS01 spectra and one KAO spectrum);
  from Gautschy-Loidl et al.~(\cite{dma4}).
}\label{f:cspec2}
\end {figure}

While the opposite (i.e. a missing feature in the synthetic spectrum
relative to the observed) is easily explained with incomplete opacity 
data, the appearance of an additional absorption feature in the 
synthetic spectra is more astonishing, and several attempts have been 
made to explain this phenomenon (cf. J{\o}rgensen et al.~\cite{jhl00}).
As discussed by Gautschy-Loidl et al.~(\cite{dma4}), however, 
dynamical models which show mass loss by stellar winds offer 
a first self-consistent explanation: 
Figure~\ref{f:cspec2} compares observed spectra of the Mira variable 
S Cep with a dynamical model that has a noticeable mass loss by a 
dust-driven stellar wind. In this case, the synthetic low-resolution 
spectra show no absorption feature around 14$\,\mu$m, since the 
photospheric absorption feature is filled in by emission from 
circumstellar material. 

This problem of the 'missing' 14 $\mu$m feature in C-rich AGB stars 
illustrates nicely how high-resolution spectra can help to clarify 
a fundamental question:
is the feature really absent (which means that there is a major 
problem with opacities and the model structures, 
cf. J{\o}rgensen et al.~\cite{jhl00}), 
or is it filled in by emission from gas layers above the 
stellar photosphere, as suggested by dynamical models ?

\subsection{High-resolution spectra - atmospheric dynamics}

High-resolution spectra of AGB stars showing the variation of 
molecular line profiles with phase are an important tool for 
studying the dynamics of the pulsating atmospheres and 
wind acceleration regions (cf. Lebzelter, this volume, 
for an overview). In particular, vibration-rotation lines 
of CO in the NIR have been used intensively for temporal 
monitoring of long-period variables (e.g. Hinkle et al.~\cite{hhr82}).
The CO molecule has several advantages over other species:
it is abundant, exists both in M- and C-type stars, forms 
at relatively high temperatures deep in the atmosphere, 
and stays chemically inert over a wide radial range. 
Lines corresponding to different transitions are formed 
in different layers, allowing to probe the velocity structure 
from the photosphere out into the wind region with just one 
molecule.

The interpretation of the observations, however, and in particular 
the derivation of absolute velocities is a non-trivial process
which requires the computation of synthetic lines from 
model structures, and comparison with the observations.
Early attempts using semi-empirical models have been mentioned 
above (Keady et al.~\cite{khr88}, Keady \& Ridgway~\cite{kr93}).
Here, we discuss a few recent examples based on self-consistent
dynamical models.

Winters et al.~(\cite{wkgs00}) studied synthetic CO fundamental and 
first overtone lines based on dust-driven wind models and compared 
these results to observations of the extreme C-rich Mira IRC+10216.
The CO fundamental lines at 4.6$\,\mu$m are formed in the zone where 
the wind has more or less reached its final outflow velocity,
showing strong P Cygni profiles and little temporal variation.
In contrast, the first overtone lines at 2.3 $\mu$m probe the 
dust formation and wind acceleration region, and the authors 
interpret the observed variations in terms of shocks and the 
formation of new dust shells. They discuss the problem of getting 
a consistent fit for the strengths and shapes of the lines 
which indicates that the mass loss rate of the model is too high 
by about a factor of 3 while the dust distribution seems consistent 
with observations.

Scholz \& Wood (\cite{sw00}) calculated synthetic CO first and 
second overtone lines and OH first overtone lines based on the 
dynamical model atmospheres by Bessell et al.~(\cite{bsw96}) and 
Hofmann et al.~(\cite{hsw98}). They derived conversion factors which 
relate velocities obtained from the Doppler profiles to the actual 
radial velocities directly above and below the shock waves in the 
models. Applying these correction factors to observed velocities 
they conclude that Mira variables are fundamental mode pulsators.

Nowotny et al. (in prep.) computed synthetic line profiles 
of CO fundamental, first and second overtone lines, as well as 
CN 0-2 red bands (at 2$\mu$m) using dynamical models of 
H{\"o}fner et al.~(\cite{dma3}). The preliminary results show 
qualitative agreement with observations, e.g., reproducing the 
S-shaped radial velocity curves observed for CO second overtone 
lines (cf. Nowotny et al., this volume)
and CN, including line doubling. The amplitudes of the 
variations obtained with the models, however, are smaller than 
the observed variations.

\section{Current projects and trends}\label{s:trends}

In the previous section we have emphasised the importance of 
self-consistency when modelling the structures and spectra of 
cool dynamical atmospheres. We have mentioned a few problems 
of current models, such as that certain micro-physical processes 
which may be relevant for obtaining a complete and realistic picture
are neglected for computational reasons.
Apart from these 'intrinsic' problems, there are also a number 
of open issues concerning parameters and boundary conditions,
as well as the overall geometry of the models.

In contrast to classical stellar atmospheres which are determined 
by three stellar parameters (e.g. mass, luminosity, effective temperature)
and the abundances of the chemical elements, dynamical models 
usually have additional parameters describing stellar pulsation
in terms of a variable inner boundary condition (see Sect.~\ref{s:dma}).
There is a lack of suitable pulsation models which could give relations 
between the basic stellar parameters and quantities describing 
the pulsation, and sometimes observed period-luminosity 
relations are used instead.
The ultimate goal, however, should be models that describe both 
the pulsation and the atmosphere and wind in a consistent way.
A first step in this direction has been taken by 
Bessell et al.~(\cite{bsw96}) and
Hofmann et al.~(\cite{hsw98}) who used pulsation models to deduce 
boundary conditions for atmospheric models. The model atmospheres 
are a 're-computation' with higher spatial resolution and more complex
micro-physics of the outermost layers of the pulsation models.

\begin{figure}[p]
\includegraphics[width=0.45\textwidth]{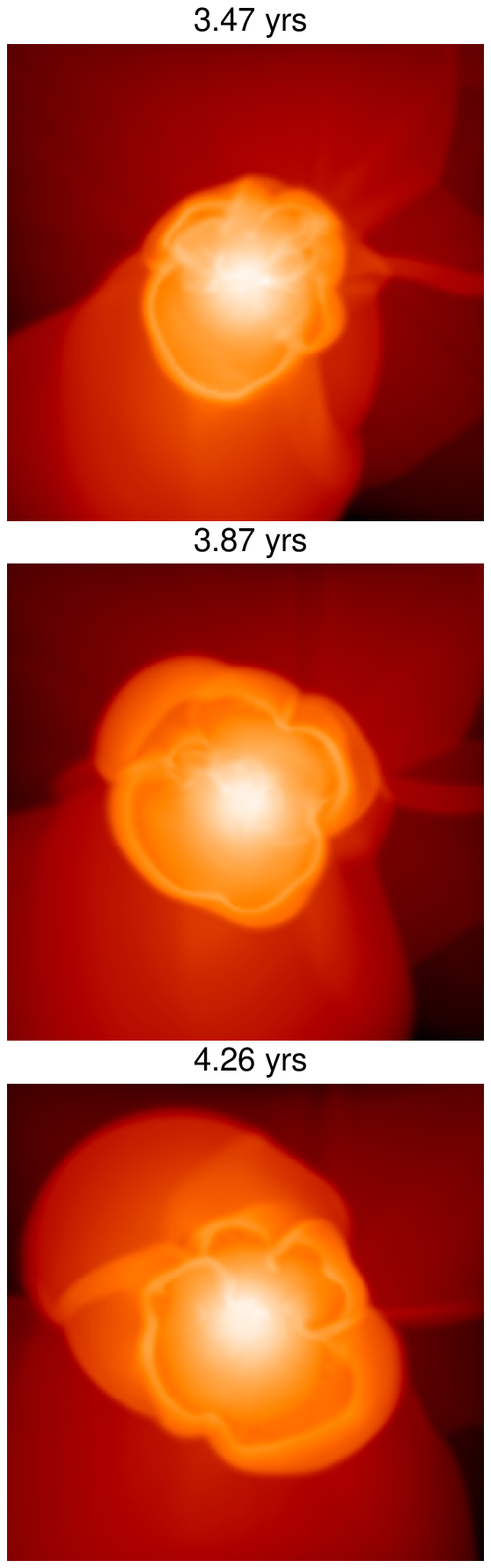}
\includegraphics[width=0.45\textwidth]{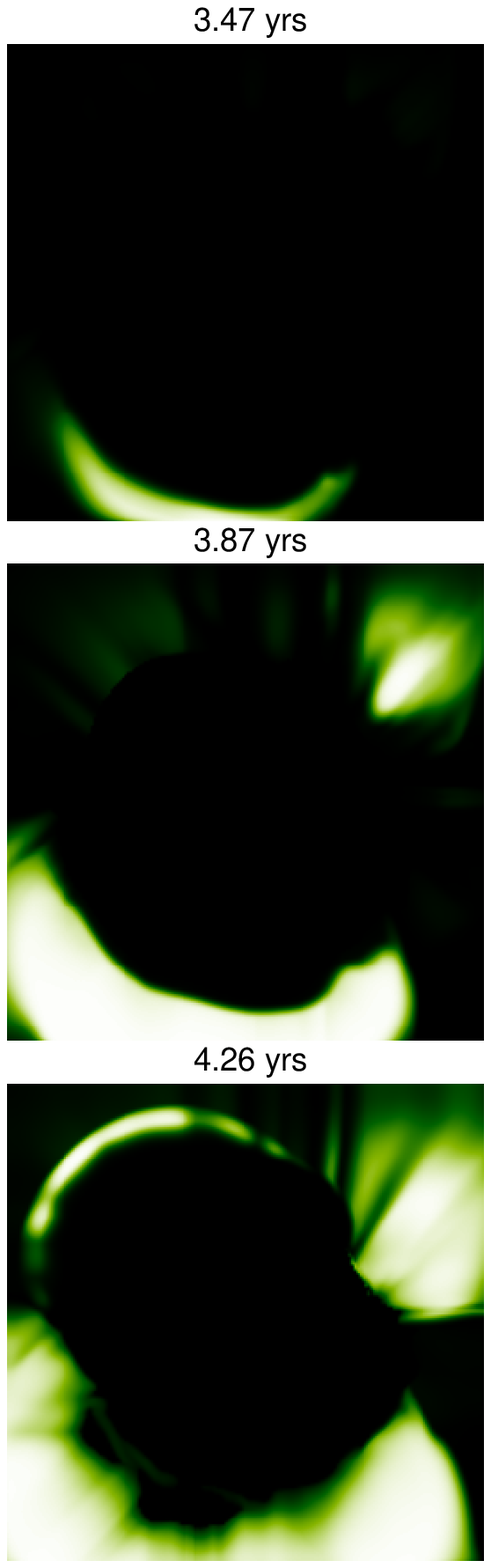}
\caption{
  Cut through the center of a 3D radiation-hydrodynamical model 
  ('star in a box') at three instants; 
  left column: gas density; right column: dust-to-gas ratio
  (dark colours indicate low values, light colours high values;
  the 'stellar surface' corresponds approximately to the strong density
  gradient contour about halfway from the center of the box,
  just above the density inversion). 
  The calculation was performed using 171x171x171 grid points.
}\label{f:3dd}
\end {figure}

A common assumption in all models discussed so far is spherical symmetry.
This assumption needs to be tested 
in the light of interferometric observations which seem to indicate 
non-spherical stellar shapes or giant spots in the photospheres of AGB stars 
(e.g. Karovska et al.~\cite{karov91}, Hofmann et al.~\cite{hbsw00}; 
see Scholz~\cite{Scholz03} for a recent review on interferometry).
Woitke et al. (\cite{wsl00}) investigated a possible instability of
dust formation in a non-homogeneous stellar atmosphere. 
This instability is caused by shadows which are cast by regions with 
a higher degree of condensation (and therefore higher optical depth),
improving the conditions for dust formation in the regions 
which are shielded from the stellar radiation. 

Large-scale convective motions are likely to cause 
inhomogeneities in the atmospheres of cool giants.
Freytag succeeded in applying three-dimensional radiation-hydrodynamical 
'star-in-a-box' simulations to supergiants (Freytag \cite{bf03}) 
and to AGB stars, demonstrating the presence of giant convections cells 
(Freytag \& H{\"o}fner \cite{fh03}). 
Recently, Freytag \& H{\"o}fner included a description of time-dependent 
dust formation into the CO$^5$BOLD code to test the influence of these 
giant convection cells on grain formation. The formation and 
growth of dust grains in atmospheres of AGB stars proceeds far from 
equilibrium, with temperature acting as a threshold and densities
of condensible material determining the efficiency of grain growth.
Therefore, one may expect that the inhomogeneities in the density 
and temperature which are caused by the convection cells will be 
imprinted on the dust distribution around the star. 
Preliminary results seem to confirm this expectation: 
Figure~\ref{f:3dd} shows a cut through the center of a 3D model 
at three instants (left column: gas density; 
right column: dust-to-gas ratio). 
The top panels show the model shortly after the equations describing 
dust formation have been switched on. The condensation process is fastest
in the dense, cool gas in the lower left corner in the wake of a shock.
Gradually, dust grains form and grow in somewhat less dense regions 
(middle panels) and in the wake of a new shock wave (upper left corner 
of the bottom panels). Note the rather sharp inner edge of the 
dust distribution due to the high temperatures close to the star.

\section{Conclusions}

Since the beginnings of time-dependent dynamical modelling more than 
two decades ago, dynamic model atmospheres for cool giant stars 
have overcome many physical and computational obstacles.
Within the last few years a transition from qualitative to 
consistent quantitative modelling is taking place.
Current models include detailed micro-physics and some 
non-equilibrium processes in the computations of 
atmospheric structures and stellar winds. 
The resulting synthetic spectra compare reasonably well 
with observations, but certain consistency issues remain
unsolved so far. This has to be kept in mind when using 
dynamical models for the interpretation of observations.

In addition to the spherical models with detailed input physics,
first prototypes of global 3D radiation-hydrodynamical models 
('star-in-a-box') have been computed recently, investigating 
the dynamics of giant convection cells and their influence 
on atmospheric structure and dust formation. These 3D models
are based on simpler micro-physics and have a lower spatial 
resolutions than the spherical models. This trade-off is 
necessary to keep computation times within acceptable limits.

In summary, the present status of dynamic model atmospheres 
and the developments that can be expected within the next few 
years look promising, regarding the interpretation of 
high-resolution IR spectra.

%INDEX%%%%%%%%%%%%%%%%%%%%%%%%%%%%%%%%%%%%%%%%%%%%%%%%%%%%%%%%%%%%%%%
% Please check with the editor of your book whether he plans to
% include a "mutual" subject index - if so, please code your entries
% in the standard syntax. For your own purposes you may print your
% "personal" index by using the following commands:
%
%\clearpage
%\addcontentsline{toc}{section}{Index}
%\flushbottom
%\printindex
%%%%%%%%%%%%%%%%%%%%%%%%%%%%%%%%%%%%%%%%%%%%%%%%%%%%%%%%%%%%%%%%%%%%%

\end{document}